\documentclass[aps,prl,preprint,groupedaddress,showpacs]{revtex4-1}

\usepackage{amsfonts}
\usepackage{graphicx}
\usepackage{epstopdf}

\newcommand{\ket}[1]{\left|#1\right\rangle}
\newcommand{\bra}[1]{\left\langle#1\right|}

\newcommand{\ie}{\textit{i.e.}}%
\newcommand{\rbfma}{\ket{\Phi^+}\bra{\Phi^+}}
\newcommand{\rbfme}{\ket{\Phi^-}\bra{\Phi^-}}
\newcommand{\rbpma}{\ket{\Psi^+}\bra{\Psi^+}}
\newcommand{\rbpme}{\ket{\Psi^-}\bra{\Psi^-}}

\begin{document}


\title{Quantum Entanglement Phase Transition in Werner State}


\author{Yuri Campbell}
\email[]{yuri.campbell@gmail.com}
\homepage[]{www.yuricampbell.org}
\author{Jos\'{e} Roberto Castilho Piqueira}
\email[]{piqueira@lac.usp.br}
\affiliation{Telecommunication and Control Engineering Department, Engineering School, University of Sao Paulo, Av. Prof. Luciano Gualberto, trv.3, n.158,\\Sao Paulo, Sao Paulo ZIP:05508-900, Brazil}


\date{\today}

\begin{abstract}
An extension to computational mechanics complexity measure is proposed in order to tackle quantum states complexity quantification.
The method is applicable to any $n-$partite state of qudits through some simple modifications.
A Werner state was considered to test this approach.
The results show that it undergoes a phase transition between entangled and separable versions of itself.
Also, results suggest interplay between quantum state complexity robustness rise and entanglement.
Finally, only via symbolic dynamics statistical analysis, the proposed method was able to distinguish separable and entangled dynamical structural differences.
\end{abstract}


\maketitle

Over the last two decades, diverse attempts to quantify complexity have been proposed, a significant amount of them make use of information-theoretic or computational tools to address this issue \cite{Shalizi2001,Kahle2009}.
Their use in various systems analysis justifies these efforts of complexity quantification in order to better understand complex systems, unraveling underlying structures and sometimes bridging together very distinct systems. 
To assess systems quantum systems there are already proposed quantum informational complexity measures \cite{Rogers2008,Mora2005,Mora2007}; however, they do not have desired features from common complexity measures proposed to classical systems because all of them are a quantum extension of Kolmogorov's algorithmic complexity \cite{kolmogorov1965three}.

Hence, they all will present the latter attribute: monotonically increasing function of disorder \cite{Shiner1999}
That is an undesirable feature for a modern complexity measure to have since its objective is to measure the degree of organization between the periodic and the random, and not how disordered a system is.
To this goal, there are a number of entropies to be used \citep{Tsallis2002}.

Therefore, a modern quantum complexity measure is a still new and unknown path to follow, by the use of quantum information theory together with modern complexity measures concepts.
Thus, in this work a well-established complexity measure is extended to accommodate quantum informational framework in order to properly tackle the quantum states complexity issue.

In this Letter, we briefly introduce computational mechanics complexity measure framework \cite{Crutchfield1989}.
Next, an iterative successive measurement procedure is defined for quantum states to adequate them to the method.
Then, the full approach is applied to a bipartite mixed state of qubits, the Werner state.
Finally, results are discussed and future work is addressed.

Originally proposed by \citet{Crutchfield1989} as a new way to quantify dynamical systems complexity, and then extended to a whole research field, computational mechanics is mainly based on a probabilistic automaton construction to imitate the analyzed system symbolic dynamics.
Through this automaton, it is possible to quantify the intrinsic computation a dynamical system performs, hence its complexity, as well.

The first step to reconstruct the automaton is a symbolic dynamics extraction through defining a state space generating partition $M_\epsilon$, constructed with cells of size $\epsilon$ which is sampled every $\tau$ time unities.

Then, using this defined measuring instrument $\{M_\epsilon,\tau\}$, a sequence of states $\{\mathbf{x}\}$ is mapped to a string of symbols $\{s,\,s\in A\}$, where $A=\{0,\cdots,k-1\}$ is the alphabet, $k\approx \epsilon^{-m_{bed}}$ is the number of partitions, and $m_{bed}$ is the data set embedded dimension.

In fact, there are a number of methods to reconstruct this automaton from the original proposition.
Other approaches were developed to reconstruct an automaton only from the observed string outputted by $\{M_\epsilon,\tau\}$.
One of these is the Causal-State Splitting Reconstruction (CSSR) algorithm \cite{Shalizi2004}.

After the automaton reconstruction, it is described with a transition matrices set $\{T^{(\gamma)}:\;\gamma\in A\}$, one for each symbol in the alphabet, defined by
$$T^{(\gamma)}\triangleq \{p_{\gamma\,,\;ij}\},$$
where $p_{\alpha\,,\;ij}$ is the probability of being in state $i$ to go to state $j$ through outputting symbol $\gamma$.
Then the probabilistic connection matrix is defined 
$$\mathbf{T}=\sum_{\{\gamma \in A\}}T^{(\gamma)}.$$

The largest eigenvalue $\lambda$ of $\mathbf{T}$ is positive \cite{Crutchfield1989}, and its associated eigenvector $\mathbf{p}=\{p_s:\,s\in \mathcal{S}\}$ has non-negative elements representing the asymptotic state probabilities, where $\mathcal{S}$ is the states set.

Finally, the $\alpha -$order automaton complexity is defined as $\mathbf{p}$ R\`{e}nyi entropy
$$C_\alpha=\frac{1}{1-\alpha}\log\sum_{s\in \mathcal{S}}p_s^\alpha,$$
which becomes $\mathbf{p}$ Shannon entropy if $\alpha\rightarrow 1$
$$C_1=-\sum_{s\in \mathcal{S}}p_s\log p_s.$$

Thus, it is clear that $C_\alpha$ is generally an intensive quantity, \ie, scale invariant; hence only automaton information fluctuation is considered.
However, if $\alpha\rightarrow 1$, $C_1$ is a extensive quantity; thus, besides automaton information fluctuation, also the number of states will influence the $C_1$ value.

Therefore, higher values of $C_1$ are a product of larger and more probabilistic equally distributed automaton.
The use of computational mechanics complexity with quantum states is not straight-forward.
It is necessary to define an iterative successive measurement process to characterize the underlying quantum state probability distributions and this will be called Quantum State Sampling (QSS).

\begin{figure}
	\centering
		\includegraphics[scale=.7]{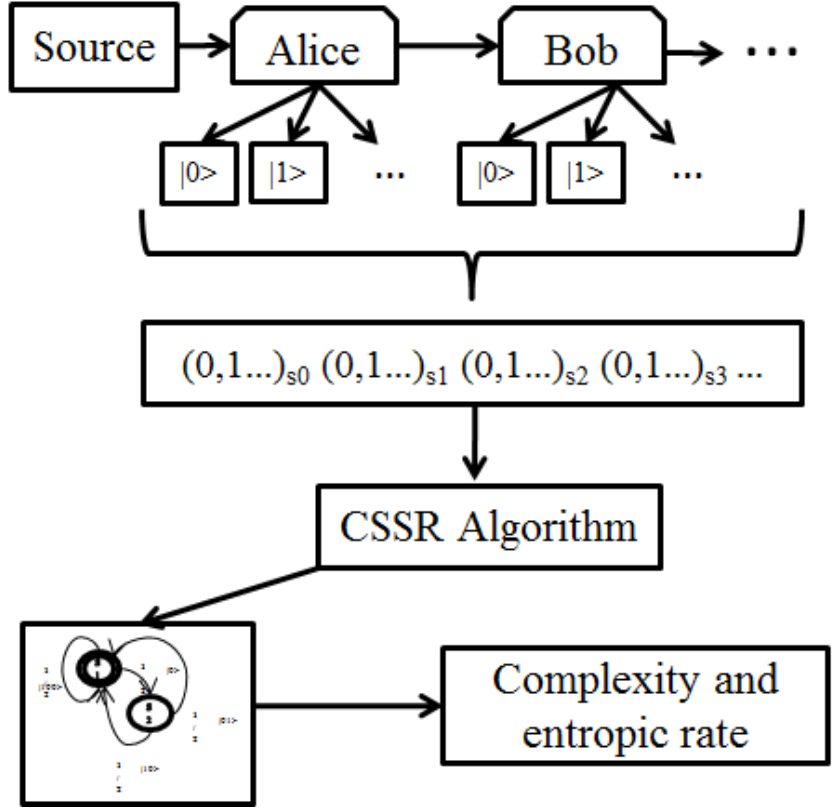}
	\caption{General quantum state sampling.}
	\label{fig:qstgen}
\end{figure}

QSS is described in Figure \ref{fig:qstgen}.
The analyzed quantum state is produced by a perfect Source.
The source generates identical quantum states and for each one of them a sequential eigenstate projective measurement procedure occurs in its sub-systems, which constitute the original quantum state.
After the last sub-system measurement takes place, the original state is completely destroyed and another one is generated by the source to undergo the same routine.
This is done iteratively, and the outcomes are generated sequentially constructing a string ($s_0s_1s_2s_3\cdots$).
Finally, this string is the input for the CSSR algorithm and therefore to computational mechanics complexity.

This approach to adapt computational mechanics complexity measure to quantum systems can be applied to any $n-$partite quantum state of qudits by a proper configuration in CSSR algorithm and maintaining this quantum state sampling process as it is.

The Werner state \cite{Vedral2006} is considered for the proposed method application
\begin{equation}
\varrho_W=\varrho_{AB}=F\rbpme +\frac{1-F}{3}\left(\rbpma+\rbfma+\rbfme\right),
\label{eq:werner}
\end{equation}
where $\ket{\Phi^\pm}$ and $\ket{\Psi^\pm}$ are the Bell states.
Thus, the sequential measurement procedure is composed by only two observers, Alice and Bob, who measure their own qubit.
But the measure outcome is outputted immediately after the measurement is made, in a sequential fashion.
This procedure produces an output string containing the outcomes of iterative measurements in perfect copies of the same state and successive in its subsystems.
Finally, this string is the input to CSSR algorithm with a string length of $1\cdot10^5$ and $L_{\text{max}}=3$, see \cite{Shalizi2004} for further details.
And then the $1-$order computational mechanics complexity is calculated through the inferred automaton.

\begin{figure}
	\centering
		\includegraphics[width=\columnwidth]{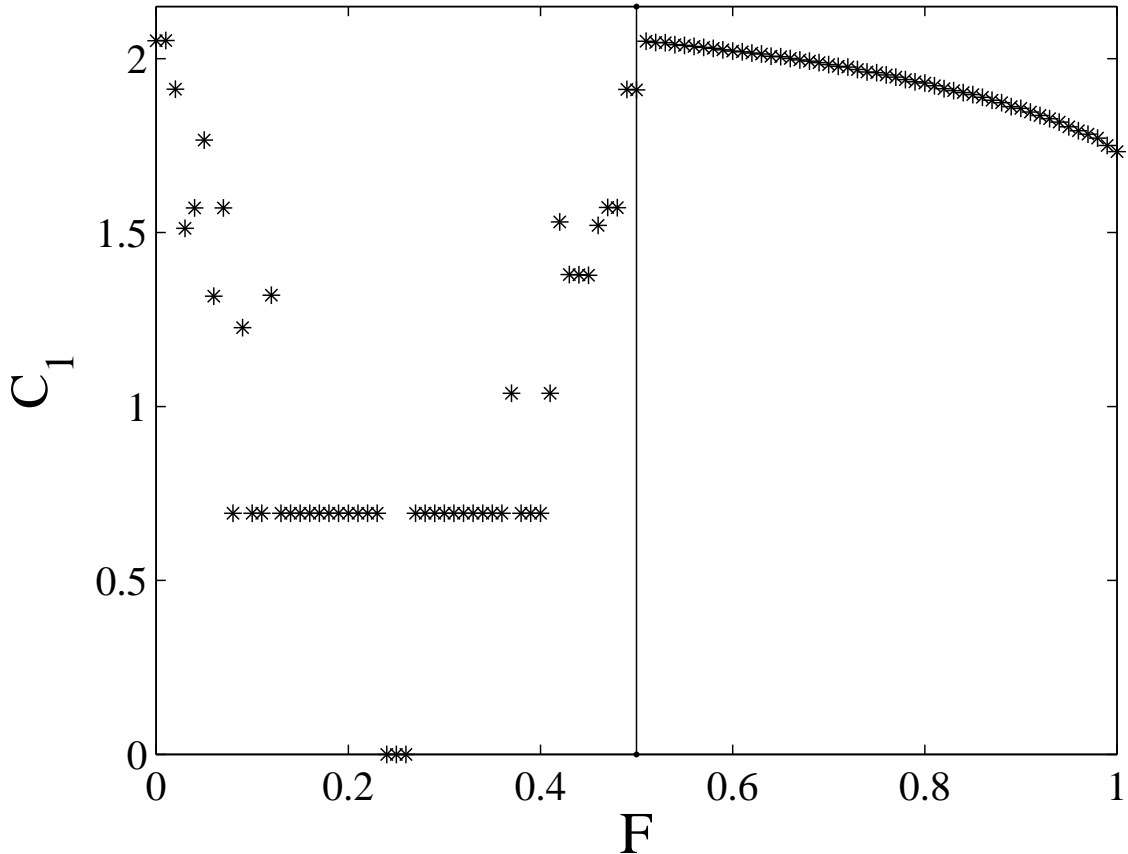}
	\caption{Werner state $C_1$ complexity.}
	\label{fig:cmbipcom}
\end{figure}

Figure \ref{fig:cmbipcom} shows complexity quantification for the Werner state as a function of its mixture parameter $F$.
It is important to point out that this state is entangled only for $F>1/2$, and maximally entangled for $F=1$.
Hence, Figure \ref{fig:cmbipcom} is split into two regions, the left one is a separable region and the right is the entangled one. 

Clearly, $C_1$ distinct behavior in these two regions strongly suggests that entanglement is a fundamental agent in quantum states complexity and responsible for a statistically significant different quantum state behavior under QSS.

\begin{figure}
	\centering
		\includegraphics[width=\columnwidth]{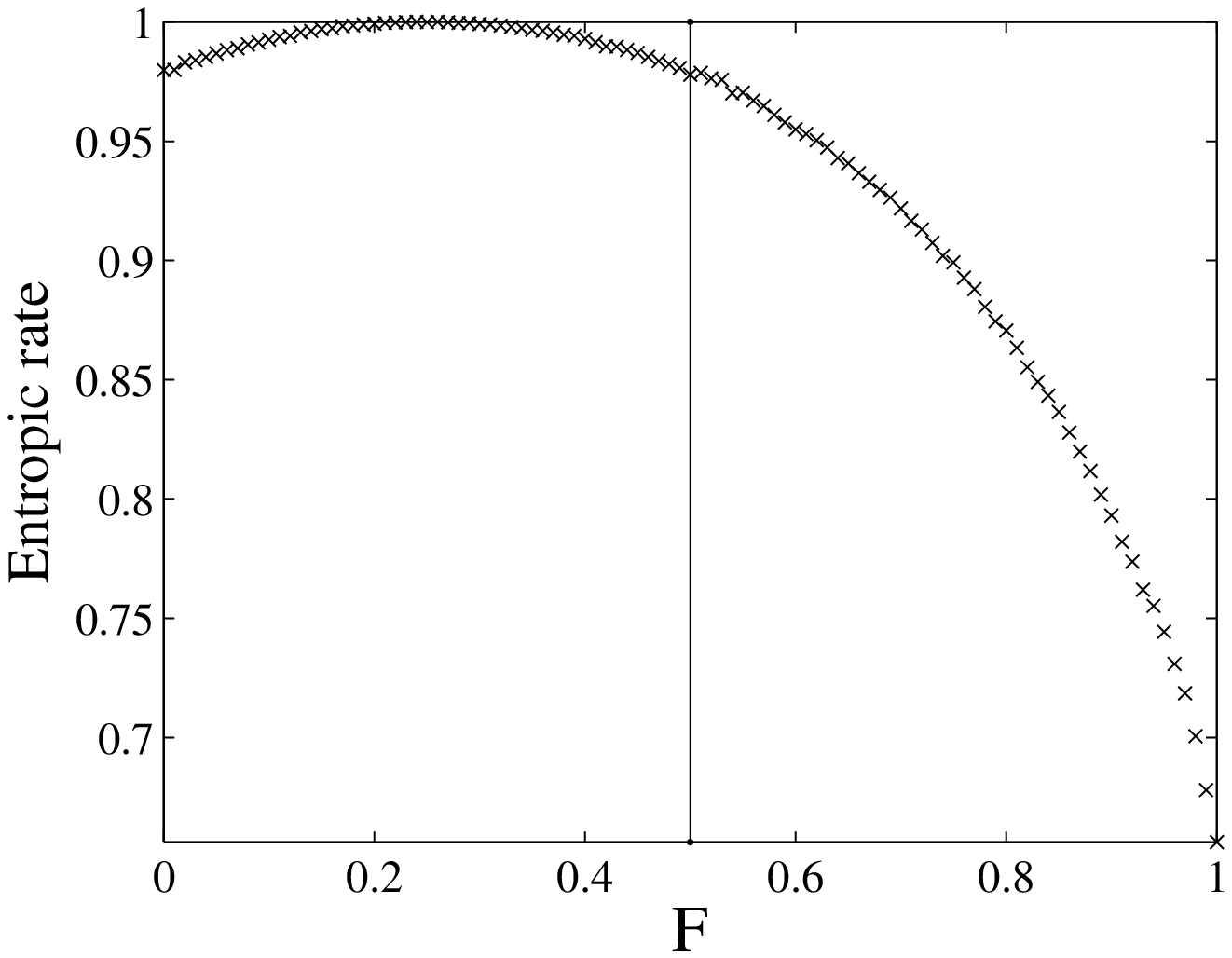}
	\caption{Werner state entropic rate.}
	\label{fig:cmbiptx}
\end{figure}

Figure \ref{fig:cmbiptx} shows the inferred automaton entropic rate in bits per symbol as a function of $F$ parameter to better analyze this effect.
Its smoothness over the full $F$ range also indicates an entanglement phenomenon capture hypothesis done by the proposed complexity quantification method.

Besides, higher complexity values are achieved at entangled-separable boundary approximation.

\begin{figure}
	\centering
		\includegraphics[width=\columnwidth]{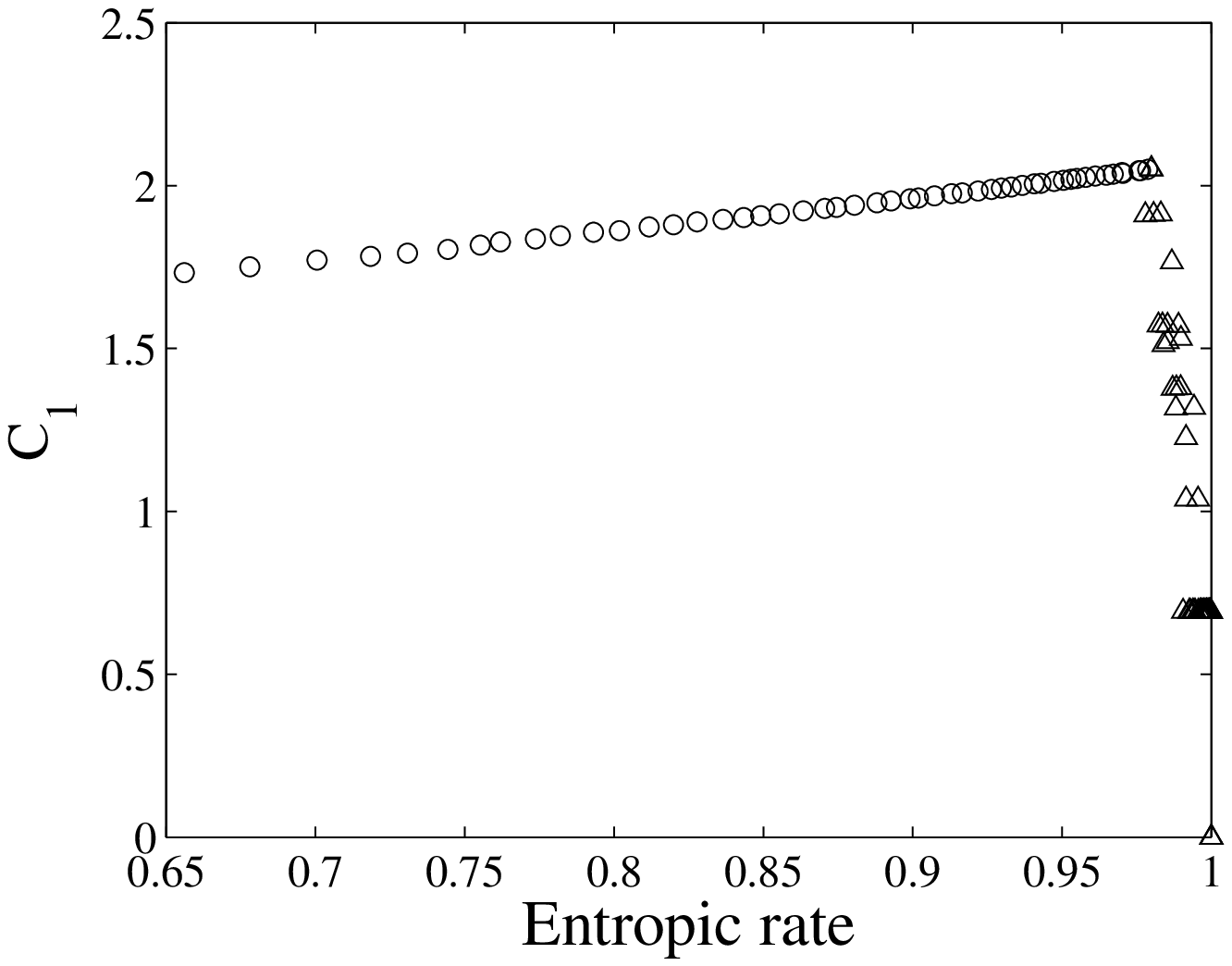}
	\caption{Werner state phase transition, the entangled states as circles and the separable as triangles.}
	\label{fig:cmbiptf}
\end{figure}

Figure \ref{fig:cmbiptf} shows $C_1$ complexity and entropic rate relation, or the Werner state complexity level as a function of system randomness.
Through this, it is easy to see in Werner state a very similar analysis to the one \citet{Crutchfield1989} made for the Logistic map.
The Werner state undergoes a phase transition exactly at the entanglement boundary, when the state goes from one region to another; its complexity behaves entirely distinctly and reaches its maximum at this transition.

Therefore, there is a system symmetry break, a structural change in its stochastic dynamics which is seized by complexity quantification.
And this change is caused by quantum entanglement.

Another appealing result is high complexity robustness inside the entangled region in Figure \ref{fig:cmbipcom}.
While for those points inside this region there is a small and smooth change in complexity, in the separable region it is abrupt and non-smooth.
That fact could point out some insight about a known question in complex quantum systems: can a quantum system complexity level be used to strengthen entanglement robustness \cite{Mora2007,Mora2005,CATERINA2006}?

To summarize, we have proposed an iterative sequential measurement procedure to quantum states called Quantum State Sampling in order to extend computational mechanics complexity measure making it also capable of analyzing these systems complexity. 
The CSSR algorithm was used as an alternative among many to extract the probabilistic automaton from system symbolic dynamics.

To test the proposed method, a Werner state was considered.
The results show that the proposed extension was capable of capturing quantum entanglement only through complexity quantification via symbolic dynamics analysis.
Also, Werner state complexity showed to be more robust inside the entangled region, suggesting interplay between these two quantities, entanglement and complexity.

Besides, a striking phenomenon was found in the Werner state complexity quantification, very similar result to the Logistic map.
A phase transition was detected in the Werner state undergoing QSS, characterizing a system symmetry break between separable and entangled versions of the Werner state.
This entanglement close relation with quantum phase transitions have already been studied in finite-dimensional case \cite{Wu2004} and infinite-dimensional also \cite{Zander2009,Rieper2010}.
And in all of these works conclusions drawn state that entanglement indicates quantum phase transitions in general \cite{Rieper2010}.

Further work needs to be done to verify those results extension for more general quantum states, at first, mixed bipartite states of qudits and mixed tripartite states of qubits.


This work is sponsored by the Brazilian National Council for Scientific and Technological Development (Conselho Nacional de Desenvolvimento Cient\'{i}fico e Tecnol\'{o}gico) - CNPq.

\bibliography{prl2}

\begin{thebibliography}{15}%
\makeatletter
\providecommand \@ifxundefined [1]{%
 \@ifx{#1\undefined}
}%
\providecommand \@ifnum [1]{%
 \ifnum #1\expandafter \@firstoftwo
 \else \expandafter \@secondoftwo
 \fi
}%
\providecommand \@ifx [1]{%
 \ifx #1\expandafter \@firstoftwo
 \else \expandafter \@secondoftwo
 \fi
}%
\providecommand \natexlab [1]{#1}%
\providecommand \enquote  [1]{``#1''}%
\providecommand \bibnamefont  [1]{#1}%
\providecommand \bibfnamefont [1]{#1}%
\providecommand \citenamefont [1]{#1}%
\providecommand \href@noop [0]{\@secondoftwo}%
\providecommand \href [0]{\begingroup \@sanitize@url \@href}%
\providecommand \@href[1]{\@@startlink{#1}\@@href}%
\providecommand \@@href[1]{\endgroup#1\@@endlink}%
\providecommand \@sanitize@url [0]{\catcode `\\12\catcode `\$12\catcode
  `\&12\catcode `\#12\catcode `\^12\catcode `\_12\catcode `\%12\relax}%
\providecommand \@@startlink[1]{}%
\providecommand \@@endlink[0]{}%
\providecommand \url  [0]{\begingroup\@sanitize@url \@url }%
\providecommand \@url [1]{\endgroup\@href {#1}{\urlprefix }}%
\providecommand \urlprefix  [0]{URL }%
\providecommand \Eprint [0]{\href }%
\providecommand \doibase [0]{http://dx.doi.org/}%
\providecommand \selectlanguage [0]{\@gobble}%
\providecommand \bibinfo  [0]{\@secondoftwo}%
\providecommand \bibfield  [0]{\@secondoftwo}%
\providecommand \translation [1]{[#1]}%
\providecommand \BibitemOpen [0]{}%
\providecommand \bibitemStop [0]{}%
\providecommand \bibitemNoStop [0]{.\EOS\space}%
\providecommand \EOS [0]{\spacefactor3000\relax}%
\providecommand \BibitemShut  [1]{\csname bibitem#1\endcsname}%
\let\auto@bib@innerbib\@empty
\bibitem [{\citenamefont {Shalizi}\ and\ \citenamefont
  {Crutchfield}(2001)}]{Shalizi2001}%
  \BibitemOpen
  \bibfield  {author} {\bibinfo {author} {\bibfnamefont {C.~R.}\ \bibnamefont
  {Shalizi}}\ and\ \bibinfo {author} {\bibfnamefont {J.~P.}\ \bibnamefont
  {Crutchfield}},\ }\href {\doibase 10.1023/A:1010388907793} {\bibfield
  {journal} {\bibinfo  {journal} {Journal of Statistical Physics}\ }\textbf
  {\bibinfo {volume} {104}},\ \bibinfo {pages} {817} (\bibinfo {year}
  {2001})}\BibitemShut {NoStop}%
\bibitem [{\citenamefont {Kahle}\ \emph {et~al.}(2009)\citenamefont {Kahle},
  \citenamefont {Olbrich}, \citenamefont {Jost},\ and\ \citenamefont
  {Ay}}]{Kahle2009}%
  \BibitemOpen
  \bibfield  {author} {\bibinfo {author} {\bibfnamefont {T.}~\bibnamefont
  {Kahle}}, \bibinfo {author} {\bibfnamefont {E.}~\bibnamefont {Olbrich}},
  \bibinfo {author} {\bibfnamefont {J.}~\bibnamefont {Jost}}, \ and\ \bibinfo
  {author} {\bibfnamefont {N.}~\bibnamefont {Ay}},\ }\href {\doibase
  10.1103/PhysRevE.79.026201} {\bibfield  {journal} {\bibinfo  {journal}
  {Physical Review E}\ }\textbf {\bibinfo {volume} {79}} (\bibinfo {year}
  {2009}),\ 10.1103/PhysRevE.79.026201}\BibitemShut {NoStop}%
\bibitem [{\citenamefont {Rogers}\ \emph {et~al.}(2008)\citenamefont {Rogers},
  \citenamefont {Vedral},\ and\ \citenamefont {Nagarajan}}]{Rogers2008}%
  \BibitemOpen
  \bibfield  {author} {\bibinfo {author} {\bibfnamefont {C.}~\bibnamefont
  {Rogers}}, \bibinfo {author} {\bibfnamefont {V.}~\bibnamefont {Vedral}}, \
  and\ \bibinfo {author} {\bibfnamefont {R.}~\bibnamefont {Nagarajan}},\ }\href
  {http://apps.isiknowledge.com/full\_record.do?product=UA\&search\_mode=GeneralSearch\&qid=73\&SID=4Ad8o8BfMFlad@cenJG\&page=1\&doc=1\&colname=WOS}
  {\bibfield  {journal} {\bibinfo  {journal} {International Journal of Quantum
  Information}\ }\textbf {\bibinfo {volume} {6}},\ \bibinfo {pages} {907}
  (\bibinfo {year} {2008})}\BibitemShut {NoStop}%
\bibitem [{\citenamefont {Mora}\ and\ \citenamefont
  {Briegel}(2005)}]{Mora2005}%
  \BibitemOpen
  \bibfield  {author} {\bibinfo {author} {\bibfnamefont {C.~E.}\ \bibnamefont
  {Mora}}\ and\ \bibinfo {author} {\bibfnamefont {H.~J.}\ \bibnamefont
  {Briegel}},\ }\href {http://www.ncbi.nlm.nih.gov/pubmed/16384044} {\bibfield
  {journal} {\bibinfo  {journal} {Physical review letters}\ }\textbf {\bibinfo
  {volume} {95}},\ \bibinfo {pages} {200503} (\bibinfo {year}
  {2005})}\BibitemShut {NoStop}%
\bibitem [{\citenamefont {Mora}\ \emph {et~al.}(2007)\citenamefont {Mora},
  \citenamefont {Briegel},\ and\ \citenamefont {Kraus}}]{Mora2007}%
  \BibitemOpen
  \bibfield  {author} {\bibinfo {author} {\bibfnamefont {C.~E.}\ \bibnamefont
  {Mora}}, \bibinfo {author} {\bibfnamefont {H.~J.}\ \bibnamefont {Briegel}}, \
  and\ \bibinfo {author} {\bibfnamefont {B.}~\bibnamefont {Kraus}},\ }\href
  {http://apps.isiknowledge.com/full\_record.do?product=UA\&search\_mode=Refine\&qid=65\&SID=4Ad8o8BfMFlad@cenJG\&page=1\&doc=6\&colname=WOS}
  {\bibfield  {journal} {\bibinfo  {journal} {International Journal of Quantum
  Information}\ }\textbf {\bibinfo {volume} {5}},\ \bibinfo {pages} {729}
  (\bibinfo {year} {2007})}\BibitemShut {NoStop}%
\bibitem [{\citenamefont {Kolmogorov}(1965)}]{kolmogorov1965three}%
  \BibitemOpen
  \bibfield  {author} {\bibinfo {author} {\bibfnamefont {A.}~\bibnamefont
  {Kolmogorov}},\ }\href {http://www.mathnet.ru/eng/ppi68} {\bibfield
  {journal} {\bibinfo  {journal} {Problemy Peredachi Informatsii}\ }\textbf
  {\bibinfo {volume} {1}},\ \bibinfo {pages} {3} (\bibinfo {year}
  {1965})}\BibitemShut {NoStop}%
\bibitem [{\citenamefont {Shiner}\ \emph {et~al.}(1999)\citenamefont {Shiner},
  \citenamefont {Davison},\ and\ \citenamefont {Landsberg}}]{Shiner1999}%
  \BibitemOpen
  \bibfield  {author} {\bibinfo {author} {\bibfnamefont {J.}~\bibnamefont
  {Shiner}}, \bibinfo {author} {\bibfnamefont {M.}~\bibnamefont {Davison}}, \
  and\ \bibinfo {author} {\bibfnamefont {P.}~\bibnamefont {Landsberg}},\ }\href
  {\doibase 10.1103/PhysRevE.59.1459} {\bibfield  {journal} {\bibinfo
  {journal} {Physical Review E}\ }\textbf {\bibinfo {volume} {59}},\ \bibinfo
  {pages} {1459} (\bibinfo {year} {1999})}\BibitemShut {NoStop}%
\bibitem [{\citenamefont {Tsallis}(2002)}]{Tsallis2002}%
  \BibitemOpen
  \bibfield  {author} {\bibinfo {author} {\bibfnamefont {C.}~\bibnamefont
  {Tsallis}},\ }\href {\doibase 10.1016/S0960-0779(01)00019-4} {\bibfield
  {journal} {\bibinfo  {journal} {Chaos, Solitons \& Fractals}\ }\textbf
  {\bibinfo {volume} {13}},\ \bibinfo {pages} {371} (\bibinfo {year}
  {2002})}\BibitemShut {NoStop}%
\bibitem [{\citenamefont {Crutchfield}\ and\ \citenamefont
  {Young}(1989)}]{Crutchfield1989}%
  \BibitemOpen
  \bibfield  {author} {\bibinfo {author} {\bibfnamefont {J.}~\bibnamefont
  {Crutchfield}}\ and\ \bibinfo {author} {\bibfnamefont {K.}~\bibnamefont
  {Young}},\ }\href {\doibase 10.1103/PhysRevLett.63.105} {\bibfield  {journal}
  {\bibinfo  {journal} {Physical Review Letters}\ }\textbf {\bibinfo {volume}
  {63}},\ \bibinfo {pages} {105} (\bibinfo {year} {1989})}\BibitemShut
  {NoStop}%
\bibitem [{\citenamefont {Shalizi}\ and\ \citenamefont
  {Klinkner}(2004)}]{Shalizi2004}%
  \BibitemOpen
  \bibfield  {author} {\bibinfo {author} {\bibfnamefont {C.~R.}\ \bibnamefont
  {Shalizi}}\ and\ \bibinfo {author} {\bibfnamefont {K.~L.}\ \bibnamefont
  {Klinkner}},\ }in\ \href {http://arxiv.org/abs/cs.LG/0406011} {\emph
  {\bibinfo {booktitle} {Uncertainty in Artificial Intelligence: Proceedings of
  the Twentieth Conference (UAI 2004)}}}\ (\bibinfo  {publisher} {AUAI Press},\
  \bibinfo {address} {Arlington, Virginia, USA},\ \bibinfo {year} {2004})\ pp.\
  \bibinfo {pages} {504----511}\BibitemShut {NoStop}%
\bibitem [{\citenamefont {Vedral}(2006)}]{Vedral2006}%
  \BibitemOpen
  \bibfield  {author} {\bibinfo {author} {\bibfnamefont {V.}~\bibnamefont
  {Vedral}},\ }\href {http://books.google.com/books?id=ffHdd6OVhGMC\&pgis=1}
  {\emph {\bibinfo {title} {{Introduction to quantum information science}}}}\
  (\bibinfo  {publisher} {Oxford University Press},\ \bibinfo {year} {2006})\
  p.\ \bibinfo {pages} {183}\BibitemShut {NoStop}%
\bibitem [{\citenamefont {Mora}\ and\ \citenamefont
  {Briegel}(2006)}]{CATERINA2006}%
  \BibitemOpen
  \bibfield  {author} {\bibinfo {author} {\bibfnamefont {C.~E.}\ \bibnamefont
  {Mora}}\ and\ \bibinfo {author} {\bibfnamefont {H.~J.}\ \bibnamefont
  {Briegel}},\ }\href
  {http://www.worldscinet.com/abstract?id=pii:S0219749906002043} {\bibfield
  {journal} {\bibinfo  {journal} {International Journal of Quantum
  Information}\ }\textbf {\bibinfo {volume} {4}},\ \bibinfo {pages} {715}
  (\bibinfo {year} {2006})}\BibitemShut {NoStop}%
\bibitem [{\citenamefont {Wu}\ \emph {et~al.}(2004)\citenamefont {Wu},
  \citenamefont {Sarandy},\ and\ \citenamefont {Lidar}}]{Wu2004}%
  \BibitemOpen
  \bibfield  {author} {\bibinfo {author} {\bibfnamefont {L.-A.}\ \bibnamefont
  {Wu}}, \bibinfo {author} {\bibfnamefont {M.}~\bibnamefont {Sarandy}}, \ and\
  \bibinfo {author} {\bibfnamefont {D.}~\bibnamefont {Lidar}},\ }\href
  {\doibase 10.1103/PhysRevLett.93.250404} {\bibfield  {journal} {\bibinfo
  {journal} {Physical Review Letters}\ }\textbf {\bibinfo {volume} {93}},\
  \bibinfo {pages} {250404} (\bibinfo {year} {2004})}\BibitemShut {NoStop}%
\bibitem [{\citenamefont {Zander}\ \emph {et~al.}(2009)\citenamefont {Zander},
  \citenamefont {Plastino},\ and\ \citenamefont {Plastino}}]{Zander2009}%
  \BibitemOpen
  \bibfield  {author} {\bibinfo {author} {\bibfnamefont {C.}~\bibnamefont
  {Zander}}, \bibinfo {author} {\bibfnamefont {A.}~\bibnamefont {Plastino}}, \
  and\ \bibinfo {author} {\bibfnamefont {A.~R.}\ \bibnamefont {Plastino}},\
  }\href@noop {} {\bibfield  {journal} {\bibinfo  {journal} {Brazilian Journal
  of Physics}\ }\textbf {\bibinfo {volume} {39}},\ \bibinfo {pages} {464}
  (\bibinfo {year} {2009})}\BibitemShut {NoStop}%
\bibitem [{\citenamefont {Rieper}\ \emph {et~al.}(2010)\citenamefont {Rieper},
  \citenamefont {Anders},\ and\ \citenamefont {Vedral}}]{Rieper2010}%
  \BibitemOpen
  \bibfield  {author} {\bibinfo {author} {\bibfnamefont {E.}~\bibnamefont
  {Rieper}}, \bibinfo {author} {\bibfnamefont {J.}~\bibnamefont {Anders}}, \
  and\ \bibinfo {author} {\bibfnamefont {V.}~\bibnamefont {Vedral}},\ }\href
  {\doibase 10.1088/1367-2630/12/2/025017} {\bibfield  {journal} {\bibinfo
  {journal} {New Journal of Physics}\ }\textbf {\bibinfo {volume} {12}},\
  \bibinfo {pages} {025017} (\bibinfo {year} {2010})}\BibitemShut {NoStop}%
\end{thebibliography}%

\end{document}